# Simulation of the response functions of Extended Range Neutron Multisphere Spectrometer using FLUKA program

WANG Panfeng(王攀峰), DING Yadong(丁亚东), WANG Qingbin(王庆斌),

MA Zhongjian(马忠剑), GUO Siming(郭思明), LI Guanjia(李冠稼)

Institute of High Energy Physics, CAS, Beijing, 100049, China

**Abstract:** In this paper, the distribution of radiation field in the CSNS spectrometer hall was simulated by FLUKA program. The results showed that the radiation field of high energy proton accelerator is dominated by neutron radiation, and the range of neutron energy is quite widely, including about eleven orders of magnitude. Nextly, work of simulation and calculation of the response functions of four Bonner spheres with simplified model is done by FLUKA and MCNP codes respectively, and the acceptable difference proved the feasibility of FLUKA program in this application and the correctness of the relevant calculation method. Through adopting the actual model, we simulate and calculate the energy response functions of Bonner sphere detectors with polyethylene layers of different diameters, including the detectors with lead layers by using FLUKA code. According to the simulation results, we select eleven detectors as the basic structure of ERNMS.

## 1 Introduction

At present three large accelerators are in operation or under construction in Institute of High Energy Physics, which are Beijing Electron-Positron Collider, high energy proton accelerator of CSNS and high current proton linac of ADS. The radiation field produced by high energy accelerator is complicated which is composed of several components such as neutron，photon, neutrino and so on. The field is dominated by neutron radiation, and the neutron energy is varying from thermal neutron to high energy of GeV magnitude. Measurement of neutron spectra plays an important reference and guidance role in the work of physical experiments and radiation protection.

The Neutron Multisphere Spectrometer which is also called Bonner Sphere Spectrometer was first introduced in 1960 by Bramblett [1] et al. A typical system consists of a series of polyethylene moderating spheres ranging from 2 inch to 12 inch in diameter with a thermal neutron sensitive detector placed at their centers.

In order to improve the accelerator protection level, and protect the health of accelerator workers and surrounding residents, radiation protection apartment of accelerator center of IHEP is working on establishing an Extended Range Neutron Multisphere Spectrometer (hereinafter referred to as ERNMS).The primary work of establishing the system is determining the structure, which can be done by reliable and accurate simulation and calculation.

In this paper, analysis of the accelerator radiation field and simulation of the response functions of the ERNMS is done by FLUKA [2] program. FLUKA is an all-particle Monte Carlo code for calculations of particle transport and interactions with matter, covering an extended range of applications spanning from accelerator shielding to target design, activation, dosimetry, detector design etc.

## 2 Analysis of the radiation field of CSNS spectrometer hall

Taking the main shielding tunnel in the CSNS spectrometer hall as an example, simulation and calculation of the radiation field around the shielding wall is done by FLUKA program.

There is a long beam tunnel of about 15 meters in the CSNS spectrometer hall. The shielding structure of the tunnel is stainless steel of 1.5 meters in thickness inside and 1 meter ordinary concrete outside. The concrete and stainless steel are all underground in order to reduce radiation dose of the workers working in the hall caused by ground scattering. Stereogram of the tunnel is edited by FLUKA program and exported by SimpleGeo, as shown in Fig. 1.

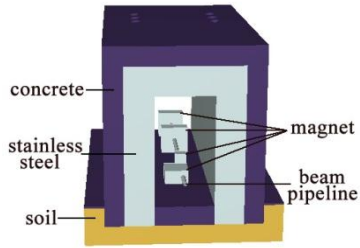

Fig. 1.　Stereogram of the tunnel

The energy of proton beam is 1.6GeV, and the beam loss rate is 1W/m. The curves of neutron and gamma dose rate with Y axis which is perpendicular to the wall are shown in Fig. 2 and Fig. 3. As we can see from the two figures, gamma dose rate is about hundredth of neutron dose rate in the tunnel, and after passing through the stainless steel and concrete layer, gamma dose rate is less than tenth of neutron dose rate at the distance of 4 meters from the beam line. So it is necessary to pay more attention to neutron dose for proton accelerators.

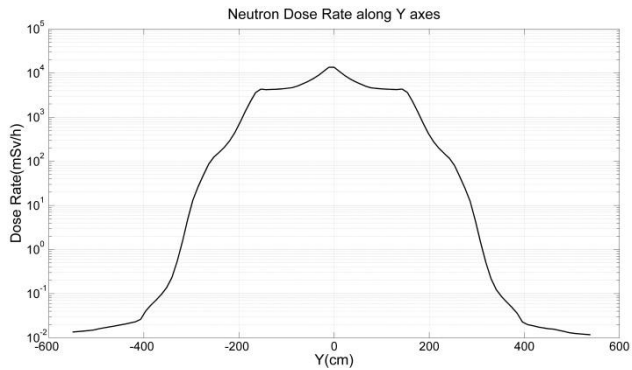 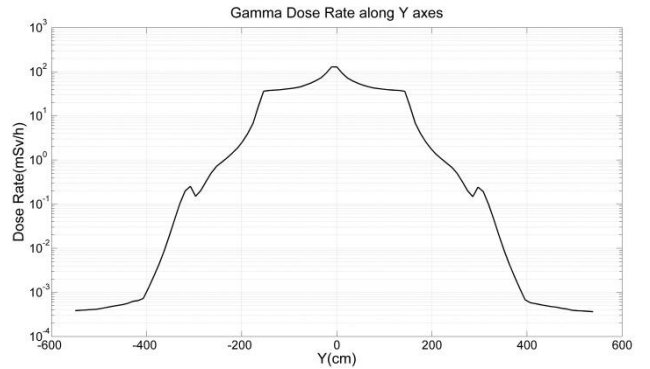

Fig. 2.　Neutron dose rate along Y axes　　　　Fig. 3. Gamma dose rate along Y axes

Fig. 4 shows neutron Lethargy spectrum of the left side of the shielding tunnel. As can be seen from the figure, not only is there a thermal peak in the neutron radiation field, but also there are two obvious peaks: evaporation peak and cascade peak at the high energy range.

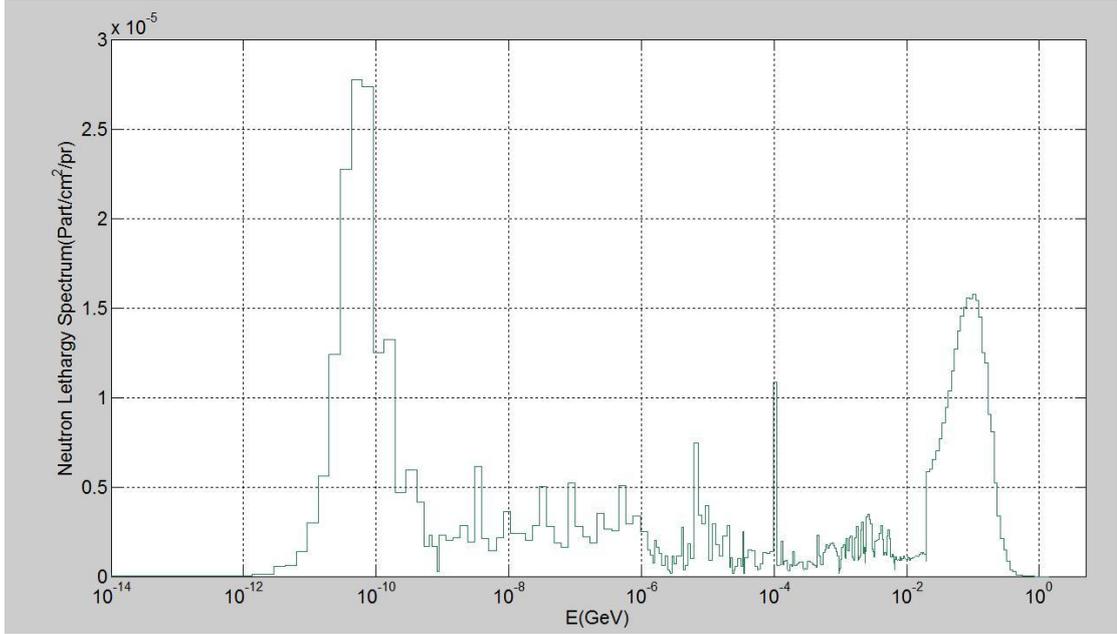

Fig. 4.   Neutron Lethargy spectrum of the left side of the shielding tunnel

In order to know the change trend of neutron flux along with energy more deeply, we divide neutron energy from thermal neutron to GeV magnitude into five groups of equal logarithmic interval, and compute the integral for neutron flux of each group. The result shows that the neutron flux whose energy is greater than 20MeV account for about 35% of the whole neutron flux.

Changing the flux integral groups to dose integral groups according to the neutron flux to dose conversion factors given by ICRP [3], we can get that the dose which is caused by neutron of energy greater than 20MeV account for 50% of the total neutron dose. This requests not only focusing on lower and middle energy neutrons, but also paying more attention to high energy neutrons when measuring neutron radiation field.

## 3  Simulation method and model
### 3.1  Operating principle

Each traditional Bonner sphere detector consists of a $^3$He-filled spherical proportional counter which is used as the central thermal-neutron-sensitive detector or bare detector and polyethylene sphere used as moderate layer. The spherical geometry generally yields an isotropic response. In order to increase the response to high energy neutrons, compensating layers such as lead were added because of their large (n,xn) cross section for high energy neutrons.

The system works on the moderate and capture method of detection. Incident neutrons interact in the polyethylene through elastic scattering with hydrogen and carbon as well as inelastic scattering off of hydrogen. As the neutrons scatter from the hydrogen and carbon they loose energy and thermalize in the material or leak out of the moderator. Some of these moderated thermal neutrons interact in the central detector via the $^3$He(n,P)T reaction.

$$^3He+n \rightarrow T+P+0.765MeV \qquad (1)$$

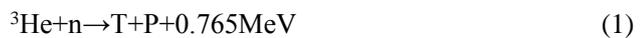

The cross section of thermal neutron with $^3$He is approximately 5400 barn which is quite large. Proton and tritium produced by the nuclear reaction ionize in the $^3$He gas. After amplifiering proportionally, the ionizing particles cause pulse charge output in the electrodes, and then recorded by nuclear electronics circuit to get counts.

The response of a Bonner sphere detector to a given neutron spectrum is given by the following equation[4]:

$$N = \int_{E_{min}}^{E_{max}} R(E)\Phi(E)\,dE \quad (2)$$

where N=counts of the detector(counts), R(E)=response function for the detector at energy E(counts per unit neutron fluence ), $\Phi(E)$=energy-dependent neutron spectrum(n/cm$^2$)

Response functions can be obtained by M-C simulation, while the counts can be done by measurement, thus we can get the neutron spectrum through appropriate unfolding methods.

### 3.2  Simulation methods

Energy response functions of neutron multisphere spectrometer are defined as the reaction counts caused by incident neutron of a unit flux from parallel source, namely as the following equation:

$$R(E) = N/\Phi(E) \quad (3)$$

where N=counts of the detector(counts), $\Phi(E)$=neutron flux at energy E(n/cm$^2$)

As to the Bonner sphere with diameter d, the response function is calculated by the following equation:

$$R_d = a_s n_{He} \sum_j \phi_j \sigma_{n,p}(\overline{E_J}) \quad (4)$$

where $a_s$=the cross sectional area of the incident neutron(cm$^2$), $n_{He}$=the atom density of $^3$He in the detector(n/cm$^3$), $\phi_j$=the neutron flux which is recorded in the $^3$He tube(n/cm$^2$), $\sigma_{n,p}(\overline{E_J})$=cross section of reaction $^3$He(n,p)T at energy $\overline{E_J} \in (E_j, E_{j+1})$(cm$^2$).

The neutron flux in the $^3$He tube was recorded by USRTRACK card of FLUKA code, while the nuclear cross section data were coming from some international neutron libraries. Response functions for each detector were calculated using 39 energy points (3 points per decade, $1.0\times10^n$, $2.0\times10^n$, $5.0\times10^n$) from 0.001eV to 5GeV.

### 3.3  Method validation

By adopting simplified model and choosing four detectors with diameters of 0(namely bare detector), 6.5, 17, 34 centimeters respectively, FLUKA and MCNP programs were used to simulate the response functions of neutron energy from 0.001eV to 20MeV. Fig. 5 shows the simulation results. As we can see from the figure, the two curves were largely consistent. Considering the difference on events of the two programs, the tiny error was acceptable. It proved that FLUKA code is feasible on simulating response functions of neutron detectors and the corresponding calculation method is correct.

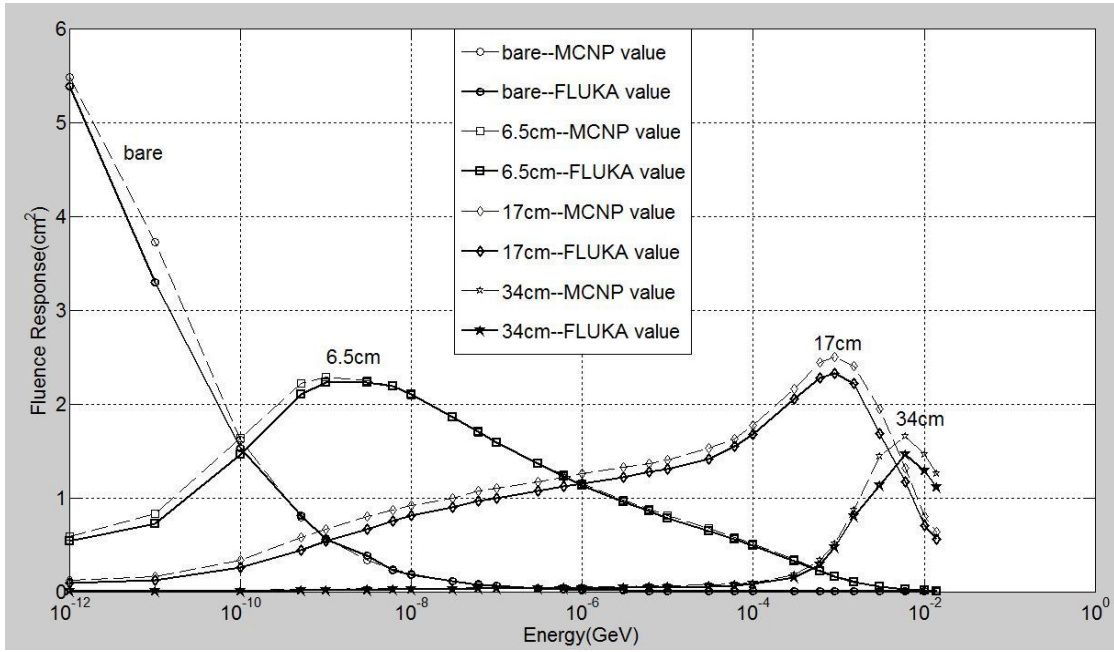

Fig. 5. The differences between FLUKA and MCNP simulation results. The dotted curve represents MCNP results, while the solid curve represents FLUKA results.

### 3.4 Simulation model and related parameters

Figure 6 shows the actual model of Bonner sphere for simulation and calculation. $^3$He-filled spherical proportional counter (type 27036, LND Ltd, USA) is used as central thermal-neutron-sensitive detector. The material of its shell is stainless steel and the inside and outside diameters are 1.96 and 2 inches. There are lead channel and related brace (called stem) at the top and bottom. The pressure of the $^3$He gas is five atmospheres and the $^3$He atom density is $12.525*10^{19}$ atoms/cm$^3$. The densities of the moderated material polyethylene and compensating material lead are 0.92 g/cm$^3$ and 11.35 g/cm$^3$ respectively. Outside the PE is air which is big enough. The outermost is blackhole which is not shown in the figure. The cross section of the incident neutron beam is circular which has the radius equal to the maximum radius of the detector. The neutrons do not to be tracked once they go into the blackhole. In order to improve accuracy, the number of incident neutrons is set to 10 million. Improper variance reduction technique will reduce the precision of the calculated results, so we don't use variance reduction technique.

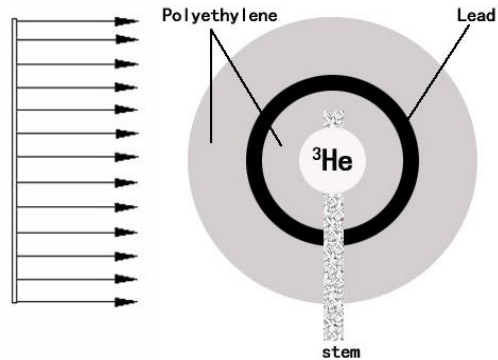

Fig. 6. The actual model of Bonner sphere for simulation of the neutron multisphere spectrometer

### 4 Simulation results and analysis

The fluence response functions of dozens of traditional Bonner detectors with different diameters are shown in Fig. 7. We can get the following conclusions from the figure: the naked detector which has no moderated and compensating materials has a significant large response for the thermal neutrons, whose peak value is about twice of other detectors' peak values; with the increase of diameter of the polyethylene layer, the peak of the curve moved from low energy to high energy, and the peaks are gathered at two areas, which are around $10^{-8}$GeV and $10^{-3}$GeV respectively; for the larger detectors, such as detectors of 15 or 18 inches, their overall response is relatively low; for all the detectors, the response at high energy is very low; the $^3$He detector which we adopt, has higher sensitivity and higher response.

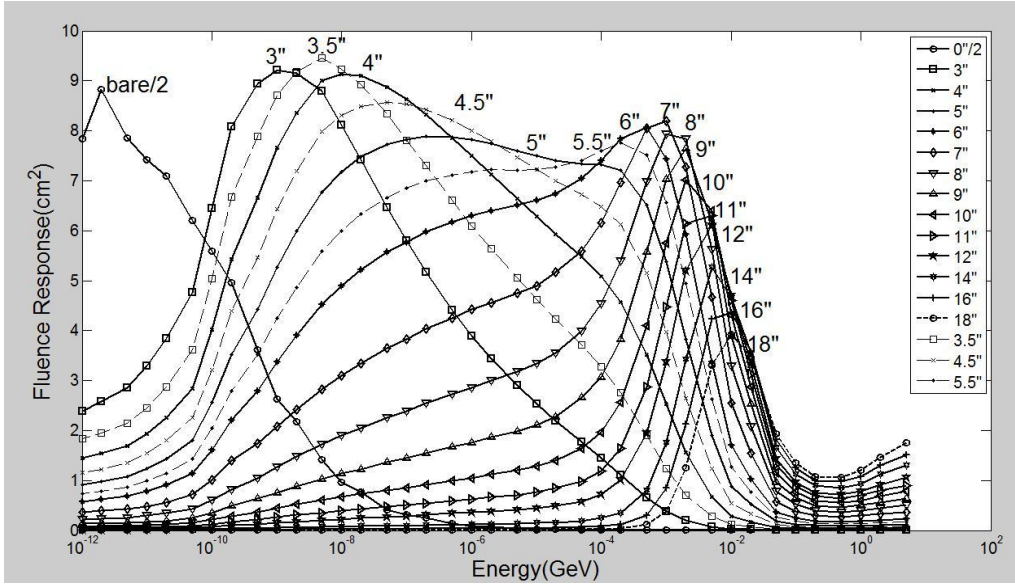

Fig. 7.  Calculated fluence response, as a function of neutron energy, for dozens of traditional Bonner detectors with different diameters

Fig. 8 shows the fluence response functions of several detectors with lead layers. As a comparison, the figure also includes the curves of the response functions of several traditional Bonner detectors. For neutron energies below 10MeV, they behave like the several intermediate sized Bonner detectors(5-8)inch, but above 20MeV the responses increase significantly. This can give additional information for unfolding procedure in high energy fields.

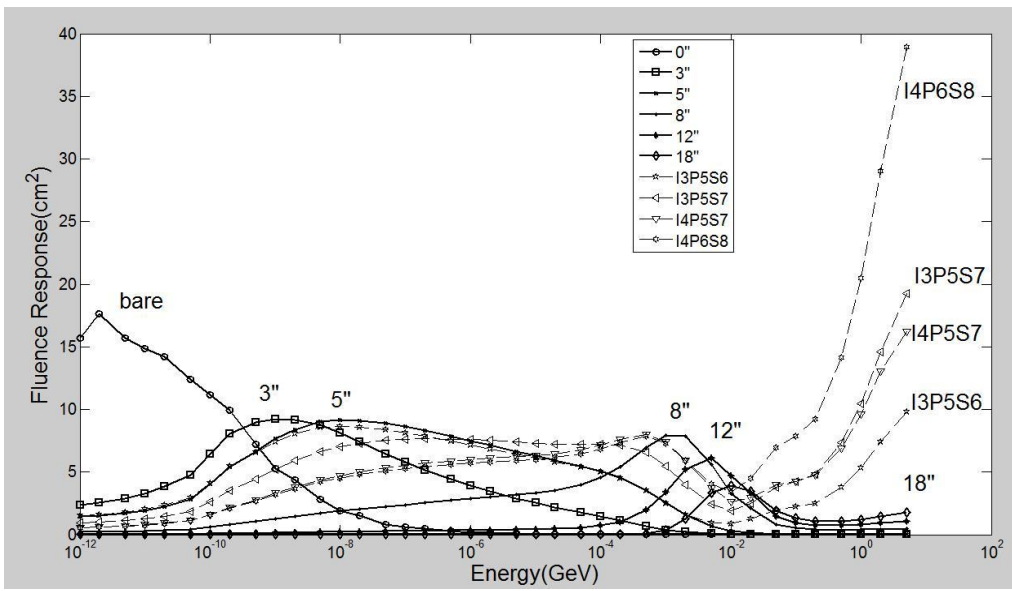

Fig. 8. Calculated fluence response, as a function of neutron energy, for several Bonner spheres and for the four detectors with lead layer. The configuration "I3P5S6": the $^3$He tube is placed inside a 3 inch PE sphere which is covered by a Pb shell with diameter of 5inch and all are imbedded in a 6 inch PE sphere.

Larger response means higher sensitivity for neutron detectors. The ideal composition of neutron multisphere spectrometer is unlimited number of detectors with different diameters. But in view of the cost and practical application, only several spheres should be selected as the fundamental structure of ERNMS. There are several selecting principles to refer to: firstly, the detectors should have larger energy response, so we should focus onto the peak value of each curve and the corresponding energy; Secondly, we must make sure that the corresponding energies of the peaks of the selected curves covered an energy range as widely as possible; thirdly, high resolution of the curves is also to be considered; last but not the least, we should grasp the balance between accuracy and applicability.

Considering the above principles and analyzing the simulation results deeply, we select nine traditional Bonner spheres with diameter 0, 3, 4, 5, 6, 7, 8, 10, 12 inch respectively and two detectors with lead layers of type I3P5S6 and I4P5S7 as the basic structure of the ERNMS system.

## 5  Conclusions

Measurement of neutron spectra has important guiding significance for the radiation protection workers. Through simulating of the CSNS spectrometer hall, we know that neutron is the most important radiation source in the radiation field of high energy accelerators，and the neutron energy covers a wide range. The main work of the paper is to simulate the response functions of dozens of Bonner spheres and several detectors with lead layers, and through analyzing of the simulation results, we select 11 different detectors as the basic structure of ERNMS.

Calculation of the response functions is the basis of developing an ERNMS system. It not only provides theory basis for determining the structure of the ERNMS, but also yields the corresponding matrix for the later unfolding procedure. Simulation results always need to be verified by factual experiments. The next work is to process, assemble and debug the spectrometer, and then do related calibration and experimental work in several standard neutron source fields and accelerator fields.

# 用FLUKA程序模拟计算扩展型中子多球谱仪的响应函数

王攀峰*，丁亚东，王庆斌，马忠剑

中国科学院高能物理研究所，北京，100049，中国

邮箱：wangpanfeng@ihep.ac.cn

**摘要**：首先使用FLUKA程序模拟计算CSNS谱仪大厅屏蔽体内外辐射场的成分，结果表明，中子在质子加速器辐射场中占主导地位，并且中子能量从热中子到GeV量级，跨越十几个量级。针对加速器辐射场的特点，本单位正研制用于测量宽能区中子能谱的中子多球谱仪。其次，分别用FLUKA和MCNP程序模拟计算多球系统简化模型的响应函数，两者的模拟结果差别很小，验证了FLUKA程序在此应用的可行性及相应计算方法的正确性。最后，采用实际模型，用FLUKA模拟计算不同直径Bonner球探测器（包括加有铅层的探测器）在单位中子注量下的响应函数，依据模拟结果，最终确定11个探测器作为扩展型中子多球谱仪系统的基本结构：。

**关键词**：中子多球谱仪、中子能谱、能量响应、FLUKA、$^3$He正比计数管